\documentclass[twocolumn,twoside,prd]{revtex4} 
\usepackage{graphicx}
\usepackage{fancyhdr}
\def \pmiss {{\,/\!\!\!p}}
\def \met {{\,/\!\!\!\!E_{T}}}
\def \jj {\,2j}

\def \jjjj {\,4j}

\begin{document}

\def \Quaero {{\sc Quaero}}
\def \pythia {{\sc Pythia}}

\title{\Quaero:  Motivation, Summary, Status}

\author{Bruce Knuteson}
\affiliation{Enrico Fermi Institute, University of Chicago}
\homepage{http://hep.uchicago.edu/~knuteson/}
\email{knuteson@fnal.gov}

\begin{abstract}
\Quaero\ is a web-based tool that automates high-$p_T$ analyses.  It has been designed with the goals of expunging exclusion contours from conference talks, obviating the necessity of ``uncorrecting'' experimental results, reducing human bias in experimental measurements, reducing by orders of magnitude the time required to perform analyses, allowing the publication of collider data in their full dimensionality, rigorously propagating systematic errors, dramatically increasing the robustness of experimental results, and facilitating the combination of results among different experiments.  \Quaero\ has been used to make a subset of D\O\ Run I data publicly available, and is being explored as a means of putting LEP data at your fingertips.  These proceedings review the motivation for \Quaero, summarize the key enabling ideas, and provide a snapshot of the project's present status.
\end{abstract}

\maketitle

\tableofcontents 

\thispagestyle{fancy}


\section{Motivation}

Current practice for testing models against collider data can be significantly improved on many fronts.

\subsection{Exclusion plots}

Take as an example the way in which the results of searches beyond the standard model are typically presented.  Starting with an enormous model space, such as the 105 parameters in the MSSM, {\em ad hoc} assumptions are imposed in order to restrict the number of free parameters to two --- this being the dimensionality of the sheet of paper on which the result will be published --- and limits are placed on the two unfixed parameters.  

\begin{figure}
\includegraphics[width=3.0in]{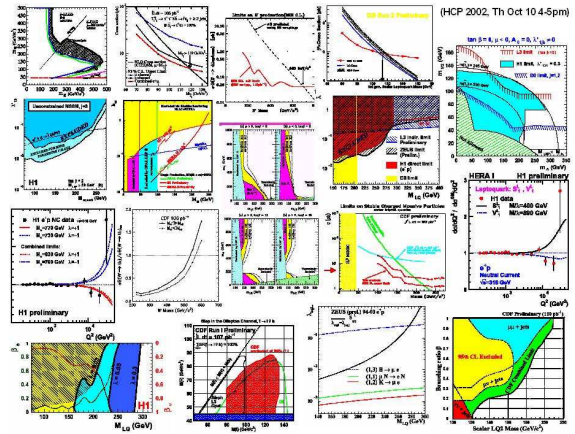}
\caption{A collage of exclusion plots shown in an hour's worth of a typical conference --- in this case Topics in Hadron Collider Physics 2002, Thursday, Oct 10, from 4-5pm~\cite{HCP2002}.}
\label{fig:ExclusionPlotCollage}
\end{figure}

Conference audiences are then inundated with the resulting exclusion plots.  The collage shown in Fig.~\ref{fig:ExclusionPlotCollage} represents an hour's worth of a typical conference --- in this case Topics in Hadron Collider Physics 2002, Thursday, Oct 10, from 4-5pm.  

Exclusion plots such as these are inherently confusing and basically useless.  They are inherently confusing because it is essentially impossible to tell exactly what model is being tested, including all assumptions that are made; in many cases this is not even clear to the author.  They are basically useless because it is nearly impossible to tell from the exclusion plot what the data have to say about some other model that happens to not lie in the two-dimensional parameter space shown~\footnote{In contrast, exclusion contours in $m_h$ for the Higgs boson search or in the plane of $\Delta m^2$ vs.~$\tan^2{\theta}$ for neutrino oscillations are extremely useful, for in these cases we really \emph{believe} the true model lies somewhere in that space.  In other cases we have no such confidence.}.

\subsection{Full dimensionality}

The results of analyses are also sometimes displayed by showing histograms of data and comparing with the predictions of several models.  This is clearly better, but limited by the fact that the data are inherently multidimensional, while histograms published in journals are inherently one- (or perhaps two-) dimensional.  Lots of information is lost in the projection.

Consider for example the simplest conceivable final state at the Fermilab Tevatron, arising from the process $p\bar{p} \rightarrow Z/\gamma^* \rightarrow e^+e^-$.  To first order, three quantities are sufficient to completely characterize each event:  these can be taken to be the invariant mass of the two electrons ($m_{ee}$), the polar angle of the positron ($\cos{\theta}$), and the transverse momentum of the $e^+e^-$ pair ($p_T^{e^+e^-}$).  No existing publication contains the three-dimensional information needed to optimally test a hypothesis against even this simple data set.  Indeed, even viewing just the three one-dimensional projections requires looking in three different publications.  In the case of CDF, see Ref.~\cite{Affolder:2001ha} for $m_{ee}$, Ref.~\cite{Affolder:2000rx} for $\cos{\theta}$, and Ref.~\cite{Affolder:1999jh} for $p_T^{e^+e^-}$.  In the case of D\O, see Ref.~\cite{Abbott:1998rr} for $m_{ee}$, Ref.~\cite{Abbott:1999yd} for $p_T^{e^+e^-}$, and let me know if you find a D\O\ publication containing the distribution of $\cos{\theta}$.

\subsection{Uncorrecting}

When publishing histograms, there is the further complication that the natural variables in which to display results are quantities measured by the experiment.  Comparison with the underlying theory, however, is facilitated if the results can be published in terms of the partons emerging from the hard scattering.  As a result, a great deal of effort is often expended in so-called ``uncorrecting'' (``unfolding,'' ``unsmearing,'' \ldots) procedures, which attempt to invert the function represented by the detector simulation.  This is a futile enterprise --- the detector simulation is a function easily and naturally understood in the forward direction in terms of a Monte Carlo propagation of particles obeying well-known laws of scattering and energy deposition, but awkwardly inverted in all but the most trivial detectors, and any uncorrecting is generally inapplicable beyond the immediate use for which it was painstakingly developed.  The natural place to compare the results of theory with experiment is in terms of the quantities observed in the detector.  

\subsection{Human bias}

\begin{figure}
\includegraphics[width=3.0in]{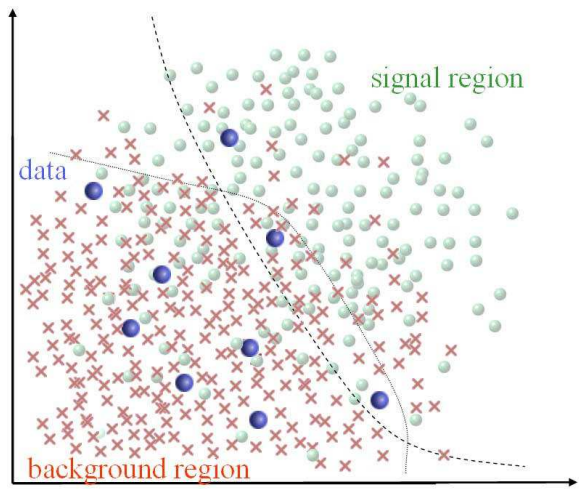}
\caption{How can a set of unbiased cuts be chosen?  In a space of two observables, simulated background events ($\times$) lie toward the lower left, simulated signal events (small balls) toward the upper right; events seen in the data are shown as large dots.  Whether the dotted or dashed contour is used to separate signal from background is subject to subtle human bias; the difference can be as simple and as seemingly innocuous as the number of nodes used in the hidden layer of a neural network.}
\label{fig:choosingCuts}
\end{figure}

Another issue deserving attention is how a set of cuts can possibly be chosen without bias.  Figure~\ref{fig:choosingCuts} shows a typical scenario, in which background populates a region in the lower left in a space of two observables, and signal populates a region in the upper right.  Simulated background events are shown as $\times$, simulated signal events are shown as small balls, and events observed in the data are shown as large balls.  Depending upon how much one believes there is signal in these data, one could choose the dashed curve (believer) or the dotted curve (disbeliever) to separate signal from background.  The difference between the two curves in this case is as simple and seemingly innocuous as the number of nodes used in the hidden layer of a neural network.

\subsection{Time}

The testing of hypotheses against data collected by large particle physics collaborations these days usually follows a rather elongated time line.  An example on the quick side of average:
\begin{itemize}
\item[]{{\bf{Jan 1, 2002}}. Theorist wakes up with a hangover and a brilliant idea.}
\item[]{{\bf{Mar 15, 2002}}. Theorist runs into a long-time experimental colleague at XXXVII Rencontres de Moriond.  The experimentalist, in a moment of weakness, decides his theoretical friend may be on to something.  He returns to his home institution, and excites his graduate student about the idea.}
\item[]{{\bf{Jun 7, 2002}}. The graduate student finishes his classes, passes his exams, and heads off to the experiment.}
\item[]{{\bf{Sep 1, 2002}}. The graduate student has mastered the experiment's analysis tools and offline framework, and plunges with gusto into the analysis.}
\item[]{{\bf{Jan 1, 2003}}. Theorist wakes up with a hangover.}
\item[]{{\bf{Jun 1, 2003}}. After overcoming various hurdles and writing ten thousand lines of code to implement a particularly clever algorithm, the student has the analysis fairly well in hand, and has obtained a preliminary result.}
\item[]{{\bf{Dec 31, 2003}}. The student's advisor being a respected and active member of the collaboration, the collaboration review process has sped through at an unprecedented clip, converging in three months.  The journal referees responded promptly and with few comments, allowing publication in the final issue of the year.}
\item[]{{\bf{Jan 1, 2004}}. Theorist wakes up, reads the article, and struggles to remember why this seemed like such a good idea.}
\end{itemize}

Most of us would really prefer something more closely resembling:
\begin{itemize}
\item[]{{\bf{11:52am}}. Physicist has an idea.}
\item[]{{\bf{11:56am}}. Physicist enters idea into his terminal.}
\item[]{{\bf{12:01pm}}. Physicist heads for lunch.}
\item[]{{\bf{12:47pm}}. Physicist receives email quantifying the extent to which the data favor (or disfavor) his idea.}
\item[]{{\bf{12:52pm}}. Physicist comes back from lunch to find results waiting for him.}
\end{itemize}
Reducing the time required to perform an analysis from two years to one hour corresponds to a speedup of over four orders of magnitude.

\subsection{Systematic errors}

A convenient scheme for assigning and propagating systematic errors in our analyses has also been lacking; the approach taken is in some cases laughably {\em ad hoc}.  This leads to the quoting of inflated systematic errors (defended as ``conservative''), resulting in a less sensitive test of the model under consideration.  Gaussian errors are nearly always assumed for convenience of calculation; any two errors are either completely correlated or completely uncorrelated (see e.g.~Ref.~\cite{Kotwal:2002uf}); propagation through anything more complicated than addition or multiplication rarely causes the student to go to the trouble of differentiating the expression to see how the results should be propagated; if he does, he is likely to get it wrong.

\subsection{Robust results}

Frequently, completely different analyses are performed for the testing of different models, even when the same subset of data is utilized.  More problematic than the inefficiency caused by this duplication of effort is the resulting difficulty of validation.  Each graduate student writes his own code for the manipulation of the data and backgrounds --- code that is used only for the purpose of one analysis, and therefore validated only to a limited extent through the obvious cross-checks that are performed in order to justify the results obtained.  Ascertaining the correctness of an analysis down to the level of potential bugs in the code thereby requires a Herculean effort on the part of the reviewing committee, which rarely spends substantial time digging through the student's C++.  The vast amounts of time typically spent optimizing a particular analysis generally decreases the robustness of the scientific conclusion, as bugs multiply most fervently in complex and clever code.  

\subsection{Combining results}

We fall down also on the combination of results, both for results from different subsets of the data within a given experiment, and for results from different experiments.  At the Tevatron in particular, there is a history of eschewing the painful process of combining results between CDF and D\O; the LEP experiments have been significantly more successful on this front.  The combination of experimental results is thus frequently performed by some theorist sitting in the back row of a conference, adding the quoted errors in quadrature (plus a little bit) and determining the combined answer.  The existence of a well-defined and well-oiled mechanism for combining results --- not only between the two Tevatron experiments, but also simultaneously with the experiments at LEP and at HERA --- would be welcome.

\subsection{Wish list}

My personal analysis wish list therefore looks something like the following~\footnote{This list does not address the understanding of the data and associated backgrounds, which will continue to require a great deal of work.  The focus is instead on how to most efficiently turn that understanding, once achieved, into meaningful statements about the underlying physics.}:
\begin{itemize}
\item{Expunge exclusion contours from conference talks}
\item{Obviate the necessity of ``uncorrecting''}
\item{Reduce human bias}
\item{Reduce analysis time by a factor of $10^4$}
\item{Publish data in their full dimensionality}
\item{Rigorously propagate systematic errors}
\item{Increase the robustness of results}
\item{Easily combine results among different experiments}
\item{All of this on the web}
\end{itemize}

\section{\Quaero: D\O\ Run I}

A first solution to these desiderata has been achieved, in the form of an algorithm named \Quaero\ (Latin for ``I search for, I seek'').  Using \Quaero, D\O\ has made a subset of its Run I data publicly available for use by the scientific community~\cite{QuaeroPRL:Abazov:2001ny}.  This represents the first such attempt by a high energy collider collaboration.

\subsection{Interface}

\begin{figure}
\includegraphics[width=3.5in]{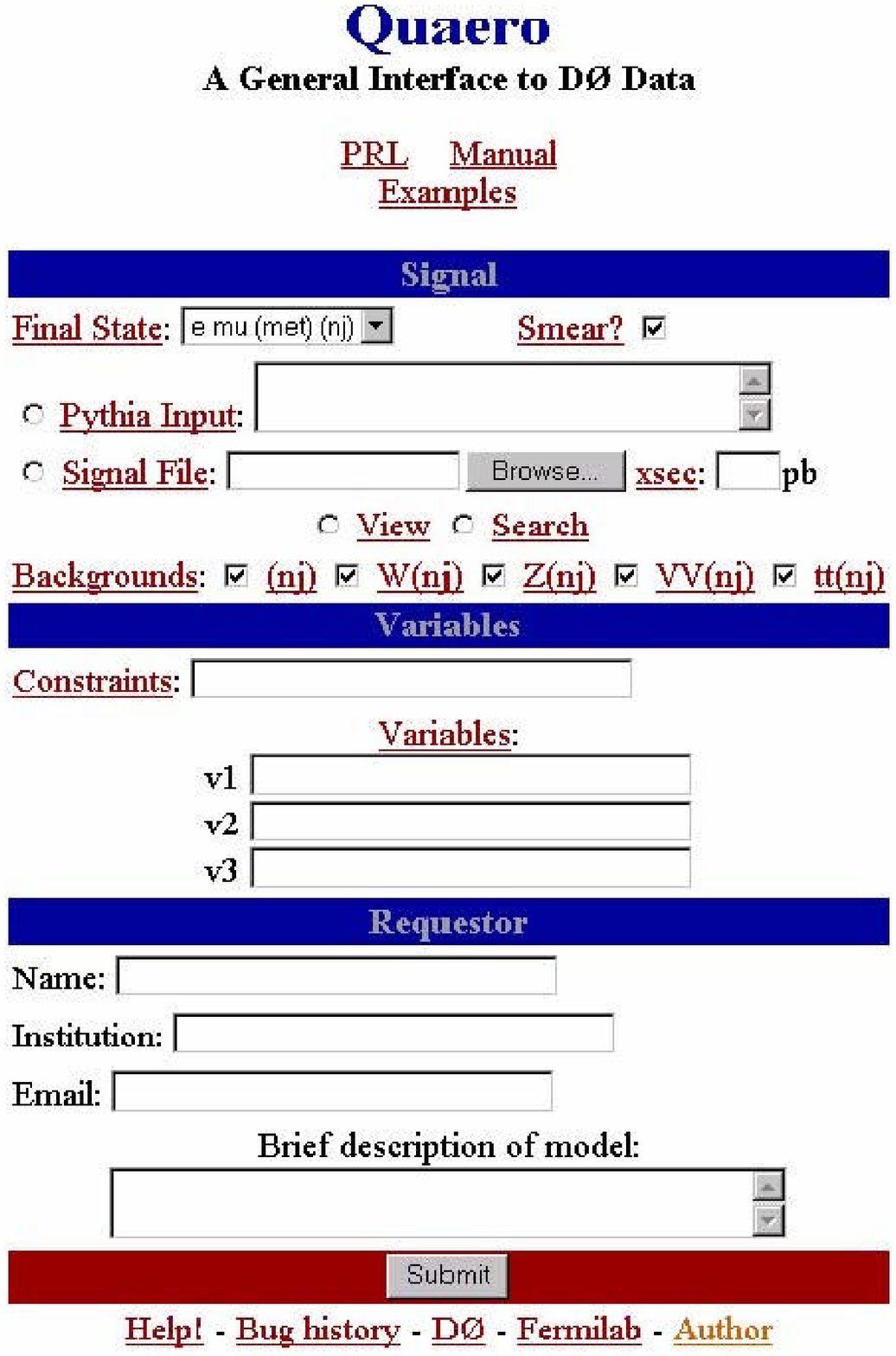}
\caption{The \Quaero\ interface for D\O\ Run I ({\tt http://quaero.fnal.gov/}).}
\label{fig:QuaeroWebPage-I}
\end{figure}

The first incarnation of \Quaero, enabling searches for new phenomena at D\O\ and the calculation of their cross sections (or limits thereon), has been available at \url{http://quaero.fnal.gov/} since the summer of 2001.  Using the web interface shown in Fig.~\ref{fig:QuaeroWebPage-I}, any high energy physicist can test a model against a subset of D\O\ Run I data, obtaining results within an hour.  

A physicist keen on a particular model begins by selecting the appropriate final state ({\underline{Final State}}) in the interface in Fig.~\ref{fig:QuaeroWebPage-I}.  The final states made available through \Quaero\ at D\O\ are those with an electron and muon; with two electrons and two or more jets; and with an electron, missing transverse energy, and two or more jets.

The physicist then provides the events predicted by the model --- either in the form of commands to \pythia~\cite{Sjostrand:2001yu} ({\underline{Pythia Input}}), which \Quaero\ will use to generate events, or in the form of a file containing the four-vectors of the events predicted ({\underline{Signal File}}), together with the cross section of the process ({\underline{xsec}}).  The physicist has the option of viewing signal, standard model background, and data ({\underline{View}}), or asking \Quaero\ to perform an optimized search for this particular signal ({\underline{Search}}).  Individual background processes can be left out of the background estimate, if desired, by unchecking the appropriate box ({\underline{Backgrounds}}).  

The physicist can provide explicit cuts ({\underline{Constraints}}), and up to three variables ({\underline{Variables}}) to distinguish signal from background.  Variables are written in a simple syntax; examples include {\tt e\_pt}, {\tt j1\_phi}, {\tt met\_pt}, {\tt mass(e1,e2)}, {\tt transversemass(e,met)}, and {\tt (j1+j2)\_pt}.
More complicated variables mixing four-vector quantities and standard C syntax, such as {\tt sqrt(j1\_phi+aplanarity())*exp(-fabs(j2\_eta))}, can also be used.  A complete description of possibilities is given in a manual available from the web page.   

After keying in his name, institution, the email to which the result should be sent, and a brief description of the model to be tested, the physicist hits the {\underline{Submit}} button, and heads for lunch.

\subsection{Algorithm}

\Quaero\ takes the events the physicist has provided (generating them, if given in the form of \pythia\ commands), runs them through a parameterized simulation of the D\O\ detector, and retains those that land in the desired final state, correctly accounting for the efficiency of object identification and the geometric acceptance of the detector.  Density estimates $p(\vec{v} | S)$ and $p(\vec{v} | B)$ are obtained for the signal $S$ and background $B$ in the variable space ${\cal V} \ni \vec{v}$ provided, and used to carve out a signal-rich region.  Within this selected region, the number of events observed in the data are compared to the number of events expected from $S$ and from $B$, and constraints on the cross section of the signal are determined.

\subsection{Results}

\begin{figure}
\includegraphics[width=3.5in]{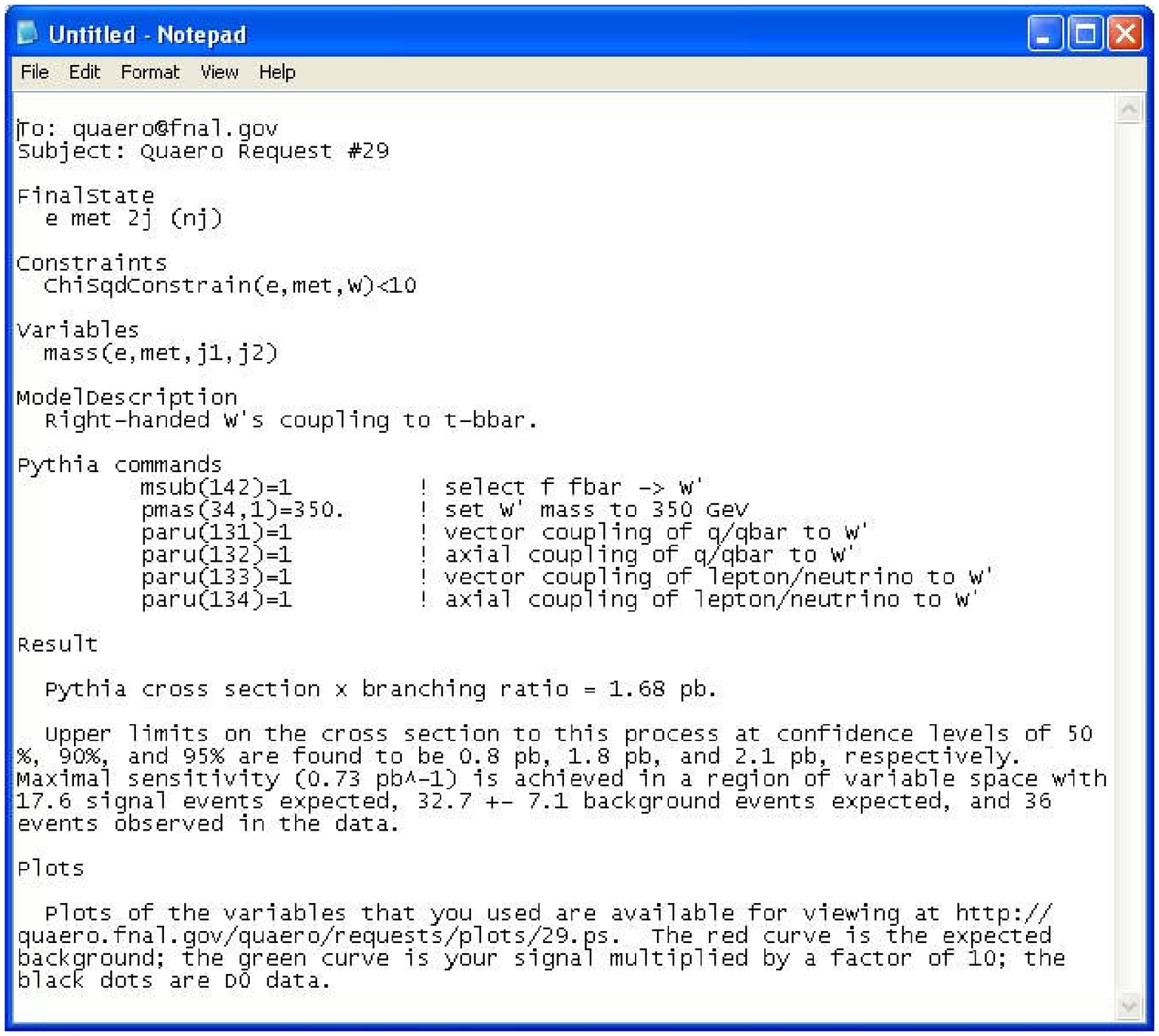}
\caption{An an example of an email returned with the result of a \Quaero\ analysis.  This was request \#29, early in the life of \Quaero; the current ticket number is over three hundred.  The requesting physicist chose the final state containing one electron, missing transverse energy, and two or more jets ({\tt e met 2j (nj)}).  The electron and inferred neutrino are constrained to a $W$ boson ({\tt ChiSqdConstrain(e,met,W)<10}), and the invariant mass of the electron, neutrino, and two leading jets ({\tt mass(e,met,j1,j2)}) is used for the purpose of distinguishing signal from background.  The signal of interest is $W_R \rightarrow t \bar{b} \rightarrow e \nu b\bar{b}$, provided to \Quaero\ in the form of commands to the \pythia\ event generator in standard notation.  No evidence for new physics is observed in this case, so \Quaero\ returns limits on the cross section of this process at various levels of confidence.  Also provided are the number of signal and background events expected in the region selected by \Quaero, the number of events observed in the data in that region, and a link to a plot of the variable used, shown in Fig.~\ref{fig:QuaeroPlotResult}.}
\label{fig:QuaeroEmailResult}
\end{figure}

\begin{figure}
\includegraphics[width=3.0in]{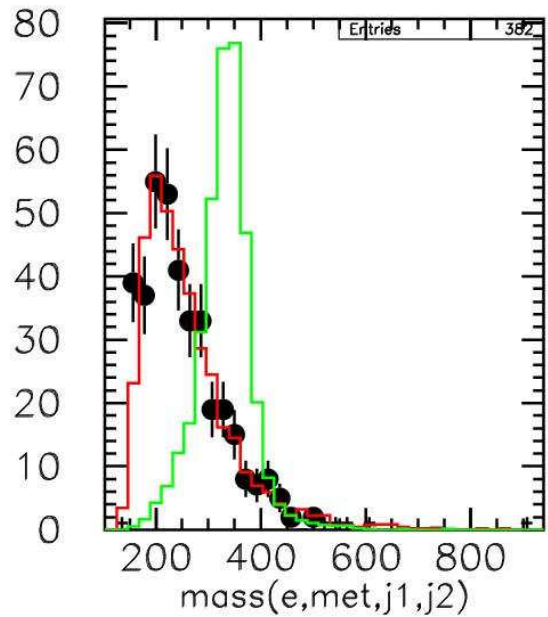}
\caption{Distribution of background (dark histogram), signal (light histogram) multiplied by a factor of 10, and D\O\ Run I data (solid points).  The variable shown is the invariant mass of the electron, inferred neutrino, and two leading jets, after constraining the electron and neutrino to a $W$ boson.  The signal peaks at the assumed mass of the $W_R$, in this case 350~GeV.}
\label{fig:QuaeroPlotResult}
\end{figure}

The resulting constraints on the signal cross section, together with plots of the variables used by \Quaero\ to determine these constraints, are returned to the physicist by email.  The email returned from an actual \Quaero\ result is shown in Fig.~\ref{fig:QuaeroEmailResult}; the plot of the variable used is shown in Fig.~\ref{fig:QuaeroPlotResult}.  

The data, backgrounds, and analysis procedure having been thoroughly reviewed by the D\O\ collaboration, the answer comes to the querying physicist as an official, D\O-approved result.  The result can be published in the querying physicist's own paper, {\it sans} D\O\ author list.  A Physical Review Letter describing \Quaero\ and its application to eleven thesis-level analyses, the results of which are provided in Table~\ref{tbl:D0RunIexamples}, has been published~\cite{QuaeroPRL:Abazov:2001ny}.  Roughly half of these analyses can be directly compared with previous CDF and D\O\ results; as expected, \Quaero\ is found to be correct and competitive in all cases.


\def \wwemubkg {19.0}
\def \wwemubkgerror {\phantom04.0}
\def \wwemusig {2.6}
\def \wwemusigeff {0.14}
\def \wwemudata {23}
\def \wwemuxseclimit {1.1\phantom0~pb}
\def \wwemuxsec {0.44}
\def \wwemuxsecLowerError {0.29}
\def \wwemuxsecUpperError {0.38}

\def \zzcbkg {19.7}
\def \zzcbkgerror {\phantom04.1}
\def \zzcsig {0.7}
\def \zzcsigeff {0.12}
\def \zzcdata {19}
\def \zzcxseclimit {0.8\phantom0~pb}
\def \zzcxsec {0.28}
\def \zzcxsecLowerError {0.21}
\def \zzcxsecUpperError {0.29}

\def \ttemetbkg {\phantom03.1}
\def \ttemetbkgerror {\phantom00.9}
\def \ttemetsig {15.3}
\def \ttemetsigeff {0.13}
\def \ttemetdata {8}
\def \ttemetxseclimit {0.8\phantom0~pb}
\def \ttemetxsec {0.39}
\def \ttemetxsecLowerError {0.19}
\def \ttemetxsecUpperError {0.21}

\def \ttemubkg {\phantom00.6}
\def \ttemubkgerror {\phantom00.2}
\def \ttemusig {1.6}
\def \ttemusigeff {0.14}
\def \ttemudata {2}
\def \ttemuxseclimit {0.4\phantom0~pb}
\def \ttemuxsec {0.14}
\def \ttemuxsecLowerError {0.08}
\def \ttemuxsecUpperError {0.15}

\def \hwwcabkg {29.6}
\def \hwwcabkgerror {\phantom06.5}
\def \hwwcasig {0.018}
\def \hwwcasigeff {0.02}
\def \hwwcadata {32}
\def \hwwcaxseclimit {11.0\phantom0~pb}
\def \hwwcaxsec {4.05}
\def \hwwcaxsecLowerError {2.75}
\def \hwwcaxsecUpperError {3.89}

\def \hwwcbbkg {66.0}
\def \hwwcbbkgerror {13.8}
\def \hwwcbsig {0.04}
\def \hwwcbsigeff {0.07}
\def \hwwcbdata {69}
\def \hwwcbxseclimit {4.4\phantom0~pb}
\def \hwwcbxsec {1.61}
\def \hwwcbxsecLowerError {1.11}
\def \hwwcbxsecUpperError {1.48}

\def \hwwccbkg {43.1}
\def \hwwccbkgerror {\phantom09.2}
\def \hwwccsig {0.02}
\def \hwwccsigeff {0.06}
\def \hwwccdata {44}
\def \hwwccxseclimit {3.6\phantom0~pb}
\def \hwwccxsec {1.26}
\def \hwwccxsecLowerError {0.87}
\def \hwwccxsecUpperError {1.21}

\def \hzzcabkg {17.9}
\def \hzzcabkgerror {\phantom03.7}
\def \hzzcasig {0.01}
\def \hzzcasigeff {0.15}
\def \hzzcadata {15}
\def \hzzcaxseclimit {0.6\phantom0~pb}
\def \hzzcaxsec {0.17}
\def \hzzcaxsecLowerError {0.12}
\def \hzzcaxsecUpperError {0.22}

\def \hzzcbbkg {18.8}
\def \hzzcbbkgerror {\phantom03.8}
\def \hzzcbsig {0.007}
\def \hzzcbsigeff {0.15}
\def \hzzcbdata {12}
\def \hzzcbxseclimit {0.4\phantom0~pb}
\def \hzzcbxsec {0.11}
\def \hzzcbxsecLowerError {0.08}
\def \hzzcbxsecUpperError {0.17}

\def \hzzccbkg {18.1}
\def \hzzccbkgerror {\phantom03.7}
\def \hzzccsig {0.006}
\def \hzzccsigeff {0.17}
\def \hzzccdata {18}
\def \hzzccxseclimit {0.6\phantom0~pb}
\def \hzzccxsec {0.2}
\def \hzzccxsecLowerError {0.15}
\def \hzzccxsecUpperError {0.2}

\def \wprimeabkg {27.7}
\def \wprimeabkgerror {\phantom06.3}
\def \wprimeasig {3.6}
\def \wprimeasigeff {0.05}
\def \wprimeadata {29}
\def \wprimeaxseclimit {3.4\phantom0~pb}
\def \wprimeaxsec {1.2}
\def \wprimeaxsecLowerError {0.82}
\def \wprimeaxsecUpperError {1.19}

\def \wprimedbkg {22.7}
\def \wprimedbkgerror {\phantom05.2}
\def \wprimedsig {3.5}
\def \wprimedsigeff {0.23}
\def \wprimeddata {27}
\def \wprimedxseclimit {0.7\phantom0~pb}
\def \wprimedxsec {0.28}
\def \wprimedxsecLowerError {0.18}
\def \wprimedxsecUpperError {0.26}

\def \wprimegbkg {\phantom02.1}
\def \wprimegbkgerror {\phantom00.8}
\def \wprimegsig {0.5}
\def \wprimegsigeff {0.26}
\def \wprimegdata {2}
\def \wprimegxseclimit {0.2\phantom0~pb}
\def \wprimegxsec {0.05}
\def \wprimegxsecLowerError {0.04}
\def \wprimegxsecUpperError {0.06}

\def \zttabkg {18.7}
\def \zttabkgerror {\phantom04.0}
\def \zttasig {13.1}
\def \zttasigeff {0.11}
\def \zttadata {20}
\def \zttaxseclimit {1.1\phantom0~pb}
\def \zttaxsec {0.39}
\def \zttaxsecLowerError {0.27}
\def \zttaxsecUpperError {0.37}

\def \zttcbkg {18.7}
\def \zttcbkgerror {\phantom04.0}
\def \zttcsig {13.7}
\def \zttcsigeff {0.12}
\def \zttcdata {20}
\def \zttcxseclimit {1.0\phantom0~pb}
\def \zttcxsec {0.37}
\def \zttcxsecLowerError {0.26}
\def \zttcxsecUpperError {0.35}

\def \zttebkg {18.7}
\def \zttebkgerror {\phantom04.0}
\def \zttesig {15.6}
\def \zttesigeff {0.14}
\def \zttedata {20}
\def \zttexseclimit {0.9\phantom0~pb}
\def \zttexsec {0.33}
\def \zttexsecLowerError {0.23}
\def \zttexsecUpperError {0.31}

\def \zttgbkg {\phantom03.8}
\def \zttgbkgerror {\phantom01.0}
\def \zttgsig {0.8}
\def \zttgsigeff {0.14}
\def \zttgdata {2}
\def \zttgxseclimit {0.3\phantom0~pb}
\def \zttgxsec {0.07}
\def \zttgxsecLowerError {0.05}
\def \zttgxsecUpperError {0.1}

\def \whbkg {37.3}
\def \whbkgerror {\phantom08.2}
\def \whsig {0.09}
\def \whsigeff {0.08}
\def \whdata {32}
\def \whxseclimit {2.0\phantom0~pb}
\def \whxsec {0.63}
\def \whxsecLowerError {0.47}
\def \whxsecUpperError {0.68}

\def \zhbkg {19.5}
\def \zhbkgerror {\phantom04.1}
\def \zhsig {0.05}
\def \zhsigeff {0.20}
\def \zhdata {25}
\def \zhxseclimit {0.8\phantom0~pb}
\def \zhxsec {0.31}
\def \zhxsecLowerError {0.19}
\def \zhxsecUpperError {0.27}

\def \lqcbkg {\phantom00.3}
\def \lqcbkgerror {\phantom00.1}
\def \lqcsig {2.8}
\def \lqcsigeff {0.33}
\def \lqcdata {0}
\def \lqcxseclimit {0.07~pb}
\def \lqcxsec {0.02}
\def \lqcxsecLowerError {0.01}
\def \lqcxsecUpperError {0.03}

\begin{table}[htb]
\centering
\begin{tabular}{clcrr}
~\\ \hline
Process & \multicolumn{1}{c}{$\epsilon_{\rm sig}$} & $\hat{b}$ & \multicolumn{1}{c}{$N_{\rm data}$} & \multicolumn{1}{c}{$\sigma^{95\%}\times {\cal B}$} \\ \hline
$WW\rightarrow e\mu\met$	& \wwemusigeff	& $\wwemubkg\pm\wwemubkgerror$	& \wwemudata	& \wwemuxseclimit  \\
$ZZ\rightarrow ee\jj$		& \zzcsigeff	& $\zzcbkg\pm\zzcbkgerror$	& \zzcdata	& \zzcxseclimit \\ 
$t\bar{t}\rightarrow e\met\jjjj$	& \ttemetsigeff	& $\ttemetbkg\pm\ttemetbkgerror$	& \ttemetdata	& \ttemetxseclimit \\
$t\bar{t}\rightarrow e\mu\met\jj$	& \ttemusigeff	& $\ttemubkg\pm\ttemubkgerror$	& \ttemudata	& \ttemuxseclimit \\ \hline
$h_{175}\rightarrow WW\rightarrow e\met\jj$	& \hwwcasigeff	& $\hwwcabkg\pm\hwwcabkgerror$	& \hwwcadata	& \hwwcaxseclimit \\
$h_{200}\rightarrow WW\rightarrow e\met\jj$	& \hwwcbsigeff	& $\hwwcbbkg\pm\hwwcbbkgerror$	& \hwwcbdata	& \hwwcbxseclimit \\
$h_{225}\rightarrow WW\rightarrow e\met\jj$	& \hwwccsigeff	& $\hwwccbkg\pm\hwwccbkgerror$	& \hwwccdata	& \hwwccxseclimit \\
$h_{200}\rightarrow ZZ\rightarrow ee\jj$	& \hzzcasigeff	& $\hzzcabkg\pm\hzzcabkgerror$	& \hzzcadata	& \hzzcaxseclimit \\
$h_{225}\rightarrow ZZ\rightarrow ee\jj$	& \hzzcbsigeff	& $\hzzcbbkg\pm\hzzcbbkgerror$	& \hzzcbdata	& \hzzcbxseclimit \\
$h_{250}\rightarrow ZZ\rightarrow ee\jj$	& \hzzccsigeff	& $\hzzccbkg\pm\hzzccbkgerror$	& \hzzccdata	& \hzzccxseclimit \\ 
$W_{200}'\rightarrow WZ\rightarrow e\met\jj$ & \wprimeasigeff & $\wprimeabkg\pm\wprimeabkgerror$ & \wprimeadata & \wprimeaxseclimit \\ 
$W_{350}'\rightarrow WZ\rightarrow e\met\jj$ & \wprimedsigeff & $\wprimedbkg\pm\wprimedbkgerror$ & \wprimeddata & \wprimedxseclimit \\ 
$W_{500}'\rightarrow WZ\rightarrow e\met\jj$ & \wprimegsigeff & $\wprimegbkg\pm\wprimegbkgerror$ & \wprimegdata & \wprimegxseclimit \\ 
$Z_{350}'\rightarrow t\bar{t}\rightarrow e\met\jjjj$	& \zttasigeff	& $\zttabkg\pm\zttabkgerror$	& \zttadata	& \zttaxseclimit \\ 
$Z_{450}'\rightarrow t\bar{t}\rightarrow e\met\jjjj$	& \zttesigeff	& $\zttebkg\pm\zttebkgerror$	& \zttedata	& \zttexseclimit \\ 
$Z_{550}'\rightarrow t\bar{t}\rightarrow e\met\jjjj$	& \zttgsigeff	& $\zttgbkg\pm\zttgbkgerror$	& \zttgdata	& \zttgxseclimit \\ \hline
$Wh_{115}\rightarrow e\met\jj$	& \whsigeff	& $\whbkg\pm\whbkgerror$	& \whdata	& \whxseclimit \\
$Zh_{115}\rightarrow ee\jj$	& \zhsigeff	& $\zhbkg\pm\zhbkgerror$	& \zhdata	& \zhxseclimit \\ 
$LQ_{225}\overline{LQ}_{225}\rightarrow ee\jj$	& \lqcsigeff	& $\lqcbkg\pm\lqcbkgerror$	& \lqcdata	& \lqcxseclimit \\ \hline
\end{tabular}
\caption{Limits on cross section $\times$ branching fraction for several thesis-level analyses performed using \Quaero.  All final states are inclusive in the number of additional jets.  The fraction of the signal sample satisfying \Quaero's selection criteria is denoted $\epsilon_{\rm sig}$; $\hat{b}$ is the number of expected background events satisfying these criteria; and $N_{\rm data}$ is the number of events in the data satisfying these criteria.  The subscripts on $h$, $W'$, $Z'$, and $LQ$ denote assumed masses, in units of GeV.  (From Ref.~\cite{QuaeroPRL:Abazov:2001ny}.)}
\label{tbl:D0RunIexamples}
\end{table}

\section{\Quaero: LEP Run II}

If a hint of new physics is revealed in Tevatron Run II, it is almost guaranteed that we will be unable to determine from the Tevatron data alone the nature of that new physics.  Unraveling the clues the Tevatron provides will require access to all high energy collider data at our disposal.  The LEP II data, in particular, may in fact be more valuable in helping us to disentangle a Tevatron hint than the rest of the Tevatron data.  If in two years a hint is observed at the Tevatron, {\tt hep-ph} will be flooded with models attempting to explain the observation.  In this event, having the ability to painlessly and quickly test many specific hypotheses against the LEP data will be invaluable.  Although there may be little more to be learned from the LEP data now, there could be a great deal to be learned from the LEP data in two years' time, illuminated by the fresh light of a Tevatron surprise.

This situation can be easily imagined.  The LEP II data set is sufficiently large that there are certainly many $3\sigma$ discrepancies.  If two models A and B attempt to explain a Tevatron hint, and Model A correctly predicts a $3\sigma$ excess in a subset of LEP data in which Model B predicts a $3\sigma$ deficit, then the LEP data favor Model A relative to Model B by a factor of roughly a million to one.  The data collected in LEP Run II therefore still have extraordinary ability to discriminate among models that may be proposed to explain hints of new physics seen at the Tevatron.

Unfortunately, the LEP II data are slowly being lost to us, as knowledgeable experimentalists move to other projects and retire.  As time goes on, the potential barrier to analyzing these data in a meaningful way grows higher and higher, not because computers fail or because Fortran becomes obsolete (it won't), but because our colleagues slowly lose their facility in the handling and understanding of these data.  The best way to ensure that the LEP II data are useful in the future is to package up --- quickly, before it is lost --- the knowledge contained in the minds of the collaborators on the four LEP experiments into an algorithm that can perform meaningful future analyses of the LEP data.

\subsection{Interface}

Improvements to the initial design of \Quaero\ are under development for LEP II and for Tevatron Run II, with potential application also to HERA I and II and the future LHC.  The new \Quaero\ is substantially more sophisticated, allowing not just the setting of cross section limits, but in fact the testing of any arbitrary model, enabling the precision measurement of parameters as well as searches for new phenomena.

\begin{figure}
\includegraphics[width=3.5in]{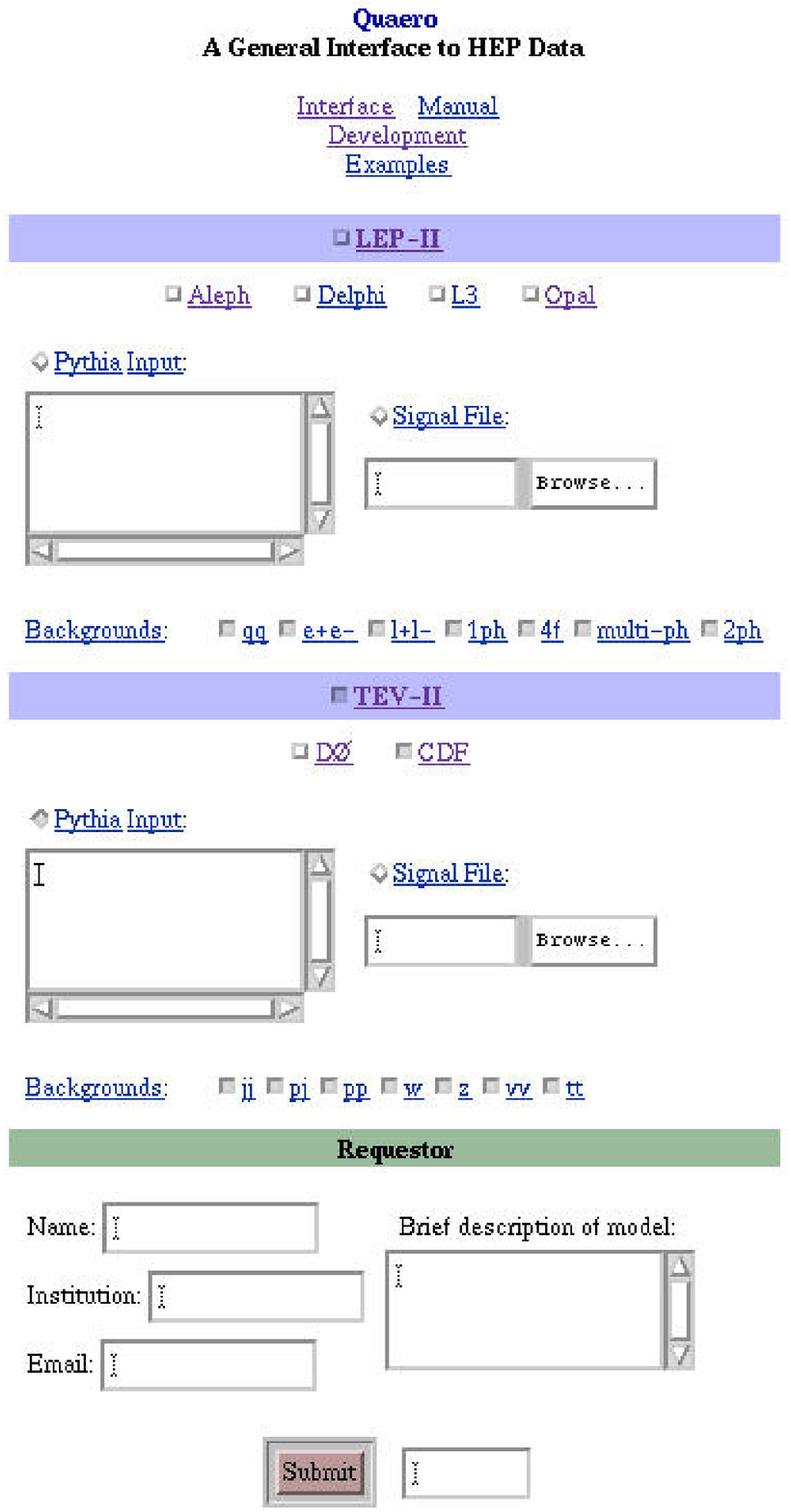}
\caption{The \Quaero\ interface under development for LEP II and Tevatron Run II.  Given a particular model, the events predicted by the model in $e^+ e^-$ collisions at $\approx 200$~GeV and in $p\bar{p}$ collisions at 1.96~TeV are provided.  These events, together with all appropriate standard model background processes, define the hypothesis to be tested.  }
\label{fig:QuaeroWebPage-II}
\end{figure}

The interface defined for LEP II and for Tevatron Run II is shown in Fig.~\ref{fig:QuaeroWebPage-II}.  The new interface is similar to that used at D\O, but with all user options removed.  The querying physicist should not need to specify the final state, explicit cuts, or useful variables --- \Quaero\ should be able to determine these itself.  Testing a particular model against collider data should be as simple as providing a model, in the form of expected events, and an email address to which the result should be sent.

\subsection{Algorithm}

The back-end interface between \Quaero\ and any experiment wishing to use \Quaero\ has also been streamlined.  An experiment needs to provide four things:
\begin{itemize}
\item[]{{\bf{Data}}. The events seen in the data, the reconstructed objects ($e^{\pm}$, $\mu^{\pm}$, $\tau^{\pm}$, $\gamma$, $\pmiss$, $j$, $b$) in those events, and the four-vectors of those objects.  An example of a data event at LEP is shown in Fig.~\ref{fig:LEPDataEvent}.}
\item[]{{\bf{Backgrounds}}. Events predicted from the standard model, the objects in those events, and the four-vectors of those objects.}
\item[]{{\bf{Systematic errors}}. Sources of systematic error, and their effect on each of the four-vector quantities.}
\item[]{{\bf{Detector simulation}}.  A simulation of the detector; this can vary from a fast parametrization to a full {\sc geant}-based simulation.}
\end{itemize}

\begin{figure}
\begin{center}{\tt
\begin{tabular}{lccc}
data &			1 &	189.2 & \\
{\bf e+} &		45.2 &	+0.11 &	0.21 \\
{\bf e-} &		47.3 &  -0.05 &	3.56 \\
{\bf b}  &		46.0 &	-0.16 &	1.71 \\
{\bf b} & 		48.2 &	-0.02 & 4.90 \\
{\bf uncl} & 		3.3 & 	+0.07 &	3.97  {\bf ;} \\
\end{tabular}}\end{center}
\caption{A sample LEP event in \Quaero\ format.  This data event has unit weight, and was collected at $\sqrt{s}=189.2$~GeV.  The event contains four reconstructed objects: a positron, an electron, and two $b$-tagged jets.  Each object is shown with its $E$ (energy, in units of GeV), $\cos{\theta}$ (cosine of the polar angle), and $\phi$ (azimuthal angle, in units of radians).  The object {\bf uncl} represents unclustered energy --- energy visible in the detector, but not clustered into any of the reconstructed objects.\label{fig:LEPDataEvent}}
\end{figure}

\Quaero\ takes the events provided to it, runs them through the detector simulation for each experiment, and partitions the resulting events into exclusive final states.  For each final state \Quaero\ constructs a low-dimensional variable space, with dimensionality limited by the number of Monte Carlo events at hand, formed from the variables in which the distributions from the standard model and the proposed model differ most.  This variable space is binned, and a likelihood ratio
\begin{equation}
{\cal L}({\cal H}) = \frac{p({\cal D}|{\cal H})}{p({\cal D}|{\text{SM}})}
\end{equation}
 is computed --- the probability of seeing the data given the proposed hypothesis ${\cal H}$ divided by the probability of seeing the data given the standard model ${\text{SM}}$.  The orthogonality of final states allows likelihoods calculated for each final state to be combined by multiplication, so that from this procedure the likelihood ratio using all relevant data is obtained.  

Systematic errors are introduced in a straightforward and intuitive way that can be made arbitrarily detailed.  For each number in each event, the effect of each source of systematic error can be assigned.  These systematic errors are then propagated into the final likelihood by numerical integration.  No Gaussian assumptions or convenient approximations need be made.  The combination of results among experiments can be handled similarly.  The way in which systematic errors are assigned lend themselves to an intuitive specification of the correlation of systematic errors among various experiments.  

We achieve at the same time significantly more robust results due to the fact that the same algorithm and code is used for all measurements.  Three hundred successful \Quaero\ analyses leads to increased confidence in the result of the three hundred first; the analogous statement does not hold if the analyses are performed by three hundred uncorrelated graduate students. 

The input to \Quaero\ is therefore simple, being just a bunch of events defining the proposed model, and the output is a single number.  A large number indicates that the model is favored relative to the standard model; a small number indicates that the model is disfavored relative to the standard model.  In addition, \Quaero\ is currently configured to return plots of the distributions of all quantities that contribute significantly to the final number returned.

An International Research Fellowship from the National Science Foundation has assisted initial efforts toward the publication of LEP data using \Quaero.  Prototypes are currently under construction within the ALEPH and L3 collaborations; policy decisions will follow the evaluation of these prototypes.

\section{Summary}

\Quaero\ is by no means a panacea.  It provides no help whatsoever in achieving a detailed understanding of instrumental features in the data or inadequacies in the detector simulation and background estimates.  It does not allow an exploration of the data for evidence of more vaguely-defined hypotheses; for this, a different algorithm is required~\cite{KnutesonThesis,SleuthPRL:Abbott:2001ke,SleuthPRD2:Abbott:2000gx,SleuthPRD1:Abbott:2000fb}.  \Quaero's sole function is to turn an existing understanding of collider data and their backgrounds into statements about the underlying physical theory by enabling efficient tests of particular hypotheses against those data.

Alternatively, \Quaero\ can be thought of as a method for publishing the results of analyses --- or the data themselves --- together with the intelligence required to make meaningful use of those data.  Physicists concerned about misuse of their data should realize that the system that has been in place now for many decades allows for easy misinterpretation of a published histogram or table of numbers by colleagues outside the collaboration lacking the detailed, requisite knowledge for making proper use of those results.  The collaboration in fact has far more control of its data if published using \Quaero, since in this case the collaboration is in complete control of how the analysis is done.  By completely eliminating all user options, the querying physicist has been rendered incapable of making a mistake.  The entire burden of the analysis rests with the collaboration, as packaged into \Quaero.

The idea for \Quaero\ began with the recognition that most high-$p_T$ analyses can be automated.  The implementation of this idea has the potential for curing --- or at least ameliorating --- several painful aspects of our field. 
\begin{itemize}
\item{By automating analyses, \Quaero\ can serve up custom exclusion plots on demand, reducing, if not eliminating, the need for these insipid plots at conferences.}
\item{\Quaero\ allows the publication of data in their full dimensionality, unlimited by the two dimensions of a sheet of paper.}
\item{\Quaero\ obviates the need for ``uncorrecting'' or ``unsmearing'' procedures by allowing input at the level of theory, but making the comparison to data at the level of what is seen in the detector.}
\item{With the high level optimization of the analysis completely prescribed by the \Quaero\ code, leaving no room for human intervention, the threat of human bias influencing the results of analyses is greatly reduced.}
\item{Automation of decisions such as the choice of variables and analysis technique reduces the time required to perform an analysis by orders of magnitude, with a corresponding savings of manpower.}
\item{The numerical propagation of systematic errors ensures a much more rigorous handling of systematics than is typically achieved, and the ability to assign the effect that each source of systematic error has on every single number in the analysis provides a much more intuitive way to think about their assignment in the first place.}
\item{The accuracy of the results obtained, while never guaranteed, is far more certain when using code that has performed sensibly in a number of previous trials than when using code validated by only a handful of individuals in the context of a single analysis.}
\item{Combining results from different final states and different experiments requires a high degree of creativity when working from the finished results; combining results with \Quaero, which performs the analysis from scratch and hence has access to all information needed to make a meaningful combination, is straightforward.}
\end{itemize}
~\\

A proof of principle of the \Quaero\ idea has been achieved, and used to make a subset of D\O\ Run I data publicly available.  An improved algorithm with much wider application is under development, and being prototyped at LEP and at the Tevatron.  

\Quaero\ has the potential for putting the LEP II data at your fingertips, on the web.  Given the tens of thousands of man years and billions of Swiss francs spent to collect these data, and the fact that we are unlikely to have $e^+e^-$ collisions at energies $\gtrsim$~200~GeV for at least another decade, the desirability of packaging the LEP data in this form should be beyond question.  This goal is kept alive in the evenings and on weekends by one person on ALEPH and one person on L3.

\begin{acknowledgments}

The D\O\ Collaboration had sufficient confidence in its data to allow their public release.  Hugh Montgomery, Mark Strovink, Ron Madaras, John Womersley, Paul Grannis, and others at D\O\ played a large role in shaping \Quaero's development.  Greg Landsberg was an invaluable co-author on the original publication.  Marcello Maggi is leading the \Quaero\ prototype effort at ALEPH; Andr\'e Holzner is leading the effort at L3.  

The development of \Quaero\ at D\O\ was supported by a Fermi/McCormick Fellowship at the University of Chicago.  The development of \Quaero\ for LEP at CERN was funded in part by an International Research Fellowship from the National Science Foundation, INT-0107322.
\end{acknowledgments}

\bibliography{chep2003}

\end{document}